\documentclass[12pt,preprint]{aastex}

\begin{document}

\title{Giant Planet Formation by Disk Instability in Low Mass Disks?}

\author{Alan P.~Boss}
\affil{Department of Terrestrial Magnetism, Carnegie Institution of
Washington, 5241 Broad Branch Road, NW, Washington, DC 20015-1305}
\email{boss@dtm.ciw.edu}

\begin{abstract}

 Forming giant planets by disk instability requires a gaseous disk that 
is massive enough to become gravitationally unstable and able to cool 
fast enough for self-gravitating clumps to form and survive. Models with 
simplified disk cooling have shown the critical importance of the ratio 
of the cooling to the orbital timescales. Uncertainties about the proper 
value of this ratio can be sidestepped by including radiative transfer.
Three-dimensional radiative hydrodynamics models of a disk with a mass of
$0.043 M_\odot$ from 4 to 20 AU in orbit around a $1 M_\odot$ protostar 
show that disk instabilities are considerably less successful in producing 
self-gravitating clumps than in a disk with twice this mass. The results are 
sensitive to the assumed initial outer disk ($T_o$) temperatures. 
Models with $T_o$ = 20 K are able to form a single self-gravitating clump,
whereas models with $T_o$ = 25 K form clumps that are not quite 
self-gravitating. These models imply that disk instability requires a disk
with a mass of at least $\sim 0.043 M_\odot$ inside 20 AU in order to form 
giant planets around solar-mass protostars with realistic disk cooling rates
and outer disk temperatures. Lower mass disks around solar-mass protostars
must rely upon core accretion to form inner giant planets.

\end{abstract}

\keywords{accretion, accretion disks --- hydrodynamics --- instabilities 
--- planets and satellites: formation --- solar system: formation}

\section{Introduction}

 The emerging census of extrasolar planets has revealed an abundance of
exoplanets, ranging from super-Earths to super-Jupiters. Mayor et al. 
(2009) suggest that $\sim$ 30\% of solar-type stars have short period
(less than 100 days) super-Earths with masses less than 30 $M_\oplus$.
The estimated frequency of giant planets with masses in the range from
0.3 to 10 $M_J$ (Jupiter-masses) inside $\sim 20$ AU is $\sim$ 10\% to 
$\sim$ 20\% (Cumming et al. 2008). Gravitational microlensing detections 
imply an even higher frequency of giant planets orbiting beyond 3 AU,
about 35\% (Gould et al. 2010). Giant planet formation thus appears to 
be a reasonably common outcome of the low-mass star formation process.

 While core accretion continues to be the most popular mechanism for 
giant planet formation (e.g., Johnson et al. 2010), disk instability seems 
to be necessary as well, at least in order to explain the formation of gas 
giant planets orbiting at great distances. HR 8799, e.g., appears to have a 
system of three giant planets, orbiting at distances of 24, 38, and 68 AU,
with masses of 10, 10, and 7 $M_J$, respectively (Marois et al. 2008).
Core accretion appears to be unable to form gas giants beyond $\sim$ 35 AU
even in the most favorable circumstances (e.g., Levison \& Stewart 2001;
Thommes, Duncan, \& Levison 2002; Chambers 2006), and gravitational scattering 
outward of planets formed closer in does not seem to lead to stable wide 
orbits (Dodson-Robinson et al. 2009; Raymond, Armitage, \& Gorelick 2010).
Disk instability appears to be the more likely mechanism for forming wide gas 
giant planets (Boss 2003, 2010; Dodson-Robinson et al. 2009; Boley 2009),
while its utility for forming planets much closer in continues to be debated
(e.g., Boss 2009).

 At a minimum, the disk instability mechanism require two conditions 
to be met in order to produce giant planets: a disk sufficiently massive
and cold enough to be gravitationally unstable, and the ability to
radiate away enough energy produced by compressional heating to allow
any clumps that form to contract toward planetary densities (e.g., Helled, 
Podolak, \& Kovetz 2006; Helled \& Bodenheimer 2010). The latter
question has been a particular focus of study, with much effort devoted
to simplified models where disk cooling occurs over a timescale $t_{cool}$. 
Gammie (2001) found that fragmentation should occur in two dimensional 
(razor-thin) disks with $\beta \le \beta_{crit} \sim 3$, 
where $\beta = t_{cool} \Omega$, with $\Omega$ being the disk's 
angular frequency. Rice et al. (2003) found that $\beta_{crit} \sim 6$ 
led to fragmentation in their three dimensional disk simulations.
Boss (2004) estimated that $\beta \sim 6$ characterized his three 
dimensional disk instability models with radiative transfer that resulted 
in clump formation.

 More recently, Meru \& Bate (2010) have performed a detailed study
of the effects of $\beta$ on disk models with varied surface density and
temperature profiles, disk masses and radii, and stellar masses, finding
that a single critical value of $\beta_{crit}$ is not always able to 
predict whether or not fragmentation occurs. In a similar vein, a
recent analysis by Nero \& Bjorkman (2009) found that their analytical
cooling time estimates were over an order of magnitude shorter than
those calculated by Rafikov (2005), and hence considerably more
supportive of fragmentation. A similar conclusion was found by Boss (2005).

 Here we completely avoid the debate over $\beta_{crit}$ by directly 
calculating disk cooling through the inclusion of radiative transfer. 
We then use this brute force approach to attack the other pre-condition 
for a disk instability leading to fragmentation, namely the disk mass. 

 Recent observations of low- and intermediate-mass pre-main-sequence stars 
imply that their disks form with masses in the range from 0.05 $M_\odot$ 
to 0.4 $M_\odot$ (Isella, Carpenter, \& Sargent 2009). These observed 
disk masses form one of the primary constraints on disk instability
models. Previous disk instability models by Boss (2002) for solar-mass 
protostars assumed disk masses of 0.091 $M_\odot$ 
from 4 to 20 AU, while those by Mayer et al. (2004) had 
disk masses ranging from 0.075 to 0.125 $M_\odot$ inside 20 AU.
We present new results here for even lower mass protoplanetary disks
(0.043 $M_\odot$), to learn if the disk instability mechanism for 
giant planet formation can continue to operate in such a low mass disk
around a solar-mass protostar.

\section{Numerical Methods}

 The calculations were performed with a numerical code that solves the 
three dimensional equations of hydrodynamics, including the energy equation, 
along with radiative transfer in the diffusion approximation and Poisson's 
equation for the gravitational potential. Compressional heating and 
radiative cooling are thus included. The same basic code has been 
used in all of the author's previous studies of disk instability. The code 
is second-order-accurate in both space and time. A complete description
of the code and of the numerous tests it passed during its development
may be found in Boss \& Myhill (1992). More recently, the radiative 
transfer solution technique has been shown to be highly accurate in 
relaxing to, and maintaining, analytical solutions for the temperature 
and radiative flux profiles for both spheres and disks of gas (Boss 2009).
Both the Jeans length (e.g., Boss et al. 2000) and the Toomre length
(Nelson 2006) criteria are monitored throughout the runs to ensure 
that any clumps that might form are not numerical artifacts. 

\section{Initial Conditions}

 The disks initially have the density distribution (Boss 1993) of
an adiabatic, self-gravitating, thick disk in near-Keplerian rotation 
about a stellar mass $M_s$

$$ \rho(R,Z)^{\gamma-1} = \rho_o(R)^{\gamma-1} - \biggl( 
{ \gamma - 1 \over \gamma } \biggr) \biggl[
\biggl( { 2 \pi G \sigma(R) \over K } \biggr) Z +
{ G M_s \over K } \biggl( { 1 \over R } - { 1 \over (R^2 + Z^2)^{1/2} }
\biggr ) \biggr], $$

\noindent where $R$ and $Z$ are cylindrical coordinates,
$\rho_o(R)$ is the midplane density, and $\sigma(R)$ is the
surface density. The adiabatic constant is $K = 1.7 \times 10^{17}$ 
(cgs units) and $\gamma = 5/3$ for the initial model; thereafter,
the disk evolves in a nonisothermal manner governed by the energy 
equation and radiative transfer (Boss \& Myhill 1992). The radial
variation of the initial midplane density is a power law that 
ensures near-Keplerian rotation throughout the disk:
$\rho_o(R) = \rho_{o4} (R_4/R)^{3/2}$,
where $\rho_{o4} = 5 \times 10^{-11}$ g cm$^{-3}$, and $R_4 = 4$ AU. 
The surface density used to define the density distribution is:
$\sigma(R) = \sigma_4 (R_4/R)^{1/2}$, where $\sigma_4 = 10^3$ g cm$^{-2}$. 
The use of this analytical surface density in the above density 
distribution results in an initial disk surface density distribution
with $\sigma \propto r^{-1/2}$ to $r^{-1}$ in the inner disk, 
steepening to $r^{-3/2}$ in the outer disk (Boss 2002). Regions where
the disk density falls to small values are considered to be in the infalling 
envelope with a density $\rho_e(R) = \rho_{e4} (R_4/R)^{3/2}$, where 
$\rho_{e4} = 10^{-14}$ g cm$^{-3}$. With $M_s = 1 M_\odot$, the disk mass 
is $M_d = 0.043 M_\odot$ from 4 to 20 AU, a mass roughly half that of 
the otherwise identical disk models in Boss (2002).

 Four different models have been computed with the above disk density 
distribution and with different combinations of outer disk temperature 
$T_{o}$ (20 and 25 K) and envelope temperature $T_{e}$ (30 and 50 K).
Model A had $T_{o}$ = 20 K and $T_{e}$ = 50 K, model B had $T_{o}$ = 25 K 
and $T_{e}$ = 50 K, model C had $T_{o}$ = 20 K and $T_{e}$ = 30 K, and
model D had $T_{o}$ = 25 K and $T_{e}$ = 30 K. The initial disk temperatures
inside 7 AU are those computed by Boss (1996) for this disk density 
distribution, yielding a midplane temperature of $T_m$ = 339 K at 4 AU and
decreasing monotonically to $T_m$ = 100 K at 7 AU; thereafter, $T_m$ 
is assumed to decrease smoothly to $T_o$ = 20 or 25 K. [In order to
err on the side of stability, the temperature is not allowed to drop 
below this initial distribution.] These choices lead to 
initial Toomre (1964) $Q$ gravitational stability criteria decreasing 
monotonically outwards from values greater than 10 inside 5 AU to minimum
$Q$ values $Q_{min}$ = 1.74 for models A and C and 1.95 for models B and D
at the outer grid boundary of 20 AU.
Higher initial $Q$ values are expected to stifle disk fragmentation, so 
models B and D are intended to test the robustness of any fragmentation
obtained in models A and C.

\section{Results}

 All four models were run initially with $N_r = 100$, $N_\theta = 45$
(effectively), $N_\phi = 256$ and $N_{Ylm} = 32$
for about 100 yr of evolution. During this time period, all four models 
evolved in a similar manner, forming multiple trailing spiral arms that
interacted with each other. The spiral arms formed throughout the disks,
but were most pronounced inside $\sim$ 10 AU. 
In models A and C, the spiral arm interactions
would occasionally lead to the formation of transient clumps. However, 
analysis of these clumps did not reveal any that were massive enough to
be considered self-gravitating and hence candidates for possible giant
planet formation. In models B and D, the spiral arms that formed were 
not as vigorous as those in models A and C, as expected given their 
slightly higher initial outer disk temperatures, and again self-gravitating
clumps did not occur. 

 After this initial phase of evolution, all four models were doubled
in their $\phi$ grid resolution and run further with $N_\phi = 512$ 
and $N_{Ylm} = 48$, effectively quadrupling the computational load
by doubling the number of grid points while halving the time step.
In order to maintain numerical stability for the energy equation solution, 
the time steps used were always small fractions of the maximum permissible 
explicit time differencing time step ($\Delta t_{CFL}$), often as small 
as 0.01 $\Delta t_{CFL}$. This resulted in painfully slow execution of the
models, each of which required approximately three years of continuous
processing on a dedicated Carnegie Alpha Cluster node.
 
 Figure 1 shows the equatorial density distribution of model A after
129 yr of evolution. Strong spiral arms are apparent from the inner
boundary at 4 AU out to $\sim$ 10 AU, as well as a number of clumps, often
still aligned with their parental spiral arms. Figure 2 depicts the
midplane temperature distribution, which rises rapidly inside $\sim$ 7 AU
to a maximum of $\sim$ 340 K at 4 AU. Comparison of Figures 1 and 2
shows that the clumps form in the region most advantageous for their
formation: just outside $\sim$ 7 AU, where the disk midplane temperatures
begin to moderate, yet as close to the center as possible, where the
orbital periods are shortest, as expected for a dynamical instability
linked to the rotation period.

 For model A at 129 yr, the maximum midplane density of $5.5 \times 
10^{-10}$ g cm$^{-3}$ occurs for the clump seen at about 6 o'clock 
in Figure 1. Figure 3 presents the midplane density and temperature 
as a function of disk radius for an azimuthal profile 
that passes through the clump at $\sim$ 6 o'clock in Figure 1.
Figure 3 shows that the maximum density occurs at a radius of $\sim$ 8 AU,
just at the radius where the temperature profile begins to rise
rapidly inward. The mass of the clump at this time is   
$\sim 0.26 M_J$, slightly above the Jeans mass of $\sim 0.24 M_J$
at the mean density ($1.1 \times 10^{-10}$ g cm$^{-3}$)
and mean temperature (26 K) of the clump. This mass estimate implies 
that the clump is self-gravitating and could be expected to
contract to higher densities if permitted by the spatial resolution
of the grid. At the radial distance of the model A clump (7 AU), $Q$ has
dropped from an initial value of 2.7 to 1.9, allowing marginal
clump formation. The tidal radius for the clump is 0.34 AU, similar
to the radial half-extent of the clump seen in Figure 3.
Note from Figure 3 that at this early phase, the clump
has not begun to undergo any significant self-heating due to 
contraction. Upward convective-like motions are present in model A, but
their vigor may not be sufficient to permit cooling on an orbital
timescale, compared to disk models with twice the disk mass, i.e.,
model HR of Boss (2004). Boss (2004) estimated an effective global
value of $\beta \sim 6$ for model HR; the reduced convective-like
motions in model A imply a value of $\beta > 6$.
   
 The clump orbits on a trajectory equivalent to a Keplerian 
orbit with a semimajor axis of 7.3 AU and an eccentricity of 0.05. 
The 6 o'clock clump shown in Figures 1, 2, and 3 first appeared
roughly 1/4 of an orbital rotation earlier, and persists for another
$\sim$ 1/2 orbital rotation before the calculation was ended after
a total of 143 yr of evolution (143 yr equals $\sim$ 19 inner orbital
rotation periods, as the disk's orbital rotation period is 7.7 yr at 
4 AU). At that final time, the estimated clump mass had increased 
slightly to $\sim 0.35 M_J$, again above the Jeans mass of 
$\sim 0.23 M_J$ at the mean density ($1.7 \times 10^{-10}$ g cm$^{-3}$)
and the slightly higher mean temperature (30 K) of the clump. 
This suggests that a protoplanet
with an initial mass of at least $\sim 0.35 M_J$ should form from this
clump. The clump's orbital eccentricity has increased to 0.09 by
this time, while its semimajor axis has decreased to 6.8 AU.

 A second distinct clump seen at about 10 o'clock in Figure 1 is not
likely to form a protoplanet, however. At a time of 129 yr, the clump's
estimated mass is $\sim 0.55 M_J$, well below its Jeans mass of 
$\sim 0.75 M_J$ at this mean density ($6.0 \times 10^{-11}$ g cm$^{-3}$)
and mean temperature (46 K). The other clumps evident in Figure 1 suffer 
from the same fate of not being massive enough to be self-gravitating.
Hence model A seems able to lead to only a single giant protoplanet.

 Figure 4 and 5 present the midplane densities and temperatures
for model B after 119 yr of evolution. Figures 4 and 5 are 
similar to Figures 1 and 2, although the spiral arms are not quite
as robust in model B as in model A, as seen in either the density
distributions of Figures 1 and 4 or the temperature distributions
of Figures 2 and 5.

 The most promising clump in model B at 119 yr occurs at 5 o'clock
in Figures 4 and 5. The estimated clump mass is $\sim 0.21 M_J$, 
below the Jeans mass of $\sim 0.35 M_J$ at the mean density 
($1.9 \times 10^{-10}$ g cm$^{-3}$) and mean temperature (40 K) 
of the clump. The clump at 7 o'clock in Figures 4 and 5 suffers
from the same problem; model B appears to be close to, but not
quite capable of forming self-gravitating clumps.

 Models C and D are identical to models A and B except for having 
envelope temperatures of 30 K instead of 50 K. While the envelope
temperature has some effect on the outcome of the evolutions,
after 137 yr model C was only able to form a single self-gravitating
clump with a mass of $\sim 0.38 M_J$ and another clump that did
not exceed the Jeans mass, as was the case for model A. Similar to 
model B, model D was unable to form a single self-gravitating clump 
after 113 yr of evolution.

\section{Discussion}

 The four models clearly show that low mass disks orbiting solar-mass
protostars are less able to form self-gravitating clumps that might
go on to form giant protoplanets than more massive disks. Boss (2002)
presented a suite of solar-mass protostar models with disk masses 
of $0.091 M_\odot$ that are otherwise much the same as the present models, 
with the exception of starting their evolutions with outer disk
temperatures ranging from 20 K to 50 K, resulting in initial minimum
Toomre $Q$ values ranging from 0.94 to 1.5. All of these 
Boss (2002) disk models formed multiple self-gravitating clumps
(see, e.g., Figure 3 of Boss 2002). The present models thus suggest
that the ability of disk instability to form self-gravitating clumps
is severely compromised as the disk mass is lowered to 
$\sim 0.043 M_\odot$. 

 The results are consistent with those obtained by Mayer et al. (2007),
who studied disks extending from 4 to 20 AU in orbit around a solar-mass
protostar using an SPH code with diffusion approximation radiative 
transfer. Mayer et al. (2007) found that when the disk mass was 
taken to be $0.05 M_\odot$, the Toomre $Q$ was below 2 in the outer
disk and strong spiral arms appeared. However, fragmentation occurred
in some of their models only when the disk mass was increased to 
$0.1-0.15 M_\odot$, with fragmentation depending on their choice of the
mean molecular weight of the disk gas and of the ability to cool from
the surface of the disk. Given that the Mayer et al. (2007) disks 
were assumed to have outer disk temperatures of 40 K, considerably
warmer than the values of 20 K and 25 K studied here, the requirement
of a disk mass higher than $0.05 M_\odot$ for fragmentation to occur 
in their models is consistent with the present models, as well as with those 
of Boss (2002), where fragmentation occurred in similar models with 
disk masses of $0.091 M_\odot$ and outer disk temperatures as high 
as 50 K.

 While envelope temperatures of 30 to 50 K appear to reasonable
bounds for a solar-mass protostar during quiescent periods 
(Chick \& Cassen 1997), the primary question arising from these
four models is what is the proper outer disk temperature?
Is $T_o =$ 20 K or 25 K beyond $\sim$ 7 AU a realistic assumption?
D'Alessio et al. (2006) presented T Tauri disk models with midplane
temperatures of $\sim$ 30 to 40 K at 10 AU, depending on the
dust grain population. Observations of the DM Tau outer disk, on scales 
though of 50 to 60 AU, imply midplane temperatures of 13 to 20 K
(Dartois, Dutrey, \& Guilloteau 2003). Observations 
of cometary ices imply disk temperatures of $\sim$ 28 K
at their formation locations (Kawakita et al. 2001). The composition
of the giant planets suggests that solids formed at 5.2 AU and beyond
at temperatures of no more than 30 to 40 K (Owens \& Encrenaz 2006). 
The present models suggest that outer disk temperatures must be as 
low as $\sim$ 20 K in order for disk instability to have a chance to 
form giant protoplanets in these relatively low mass disks, and it is 
unclear at present if such low outer disk temperatures are realistic 
or not.

\section{Conclusions}

 Boss (2002) found that robust disk instablities could occur inside 20 AU 
in disks with a mass of $\sim 0.091 M_\odot$. The present models show that 
when the disk mass inside 20 AU is halved, the ability of disk instability 
to produce viable, self-gravitating clumps is signficantly compromised,
when self-consistently-calculated disk cooling rates are employed.
Disk instability thus appears to be only a marginally effective process 
in a disk with $\sim 0.04 M_\odot$, and is unlikely to lead to giant planet
formation around solar-mass protostars with disks significantly less 
massive than $\sim 0.04 M_\odot$. Clearly core accretion remains as the 
favored formation mechanism for giant planets in such lower mass disks.

 I thank the referee for a number of perceptive comments, Sandy Keiser for 
computer systems support and John Chambers for advice on orbit determinations.
This research was supported in part by NASA Planetary Geology and Geophysics 
grant NNX07AP46G, and is contributed in part to NASA Astrobiology Institute
grant NNA09DA81A. The calculations were performed on the Carnegie Alpha
Cluster, the purchase of which was partially supported by NSF Major Research
Instrumentation grant MRI-9976645.

\begin{figure}
\plotone{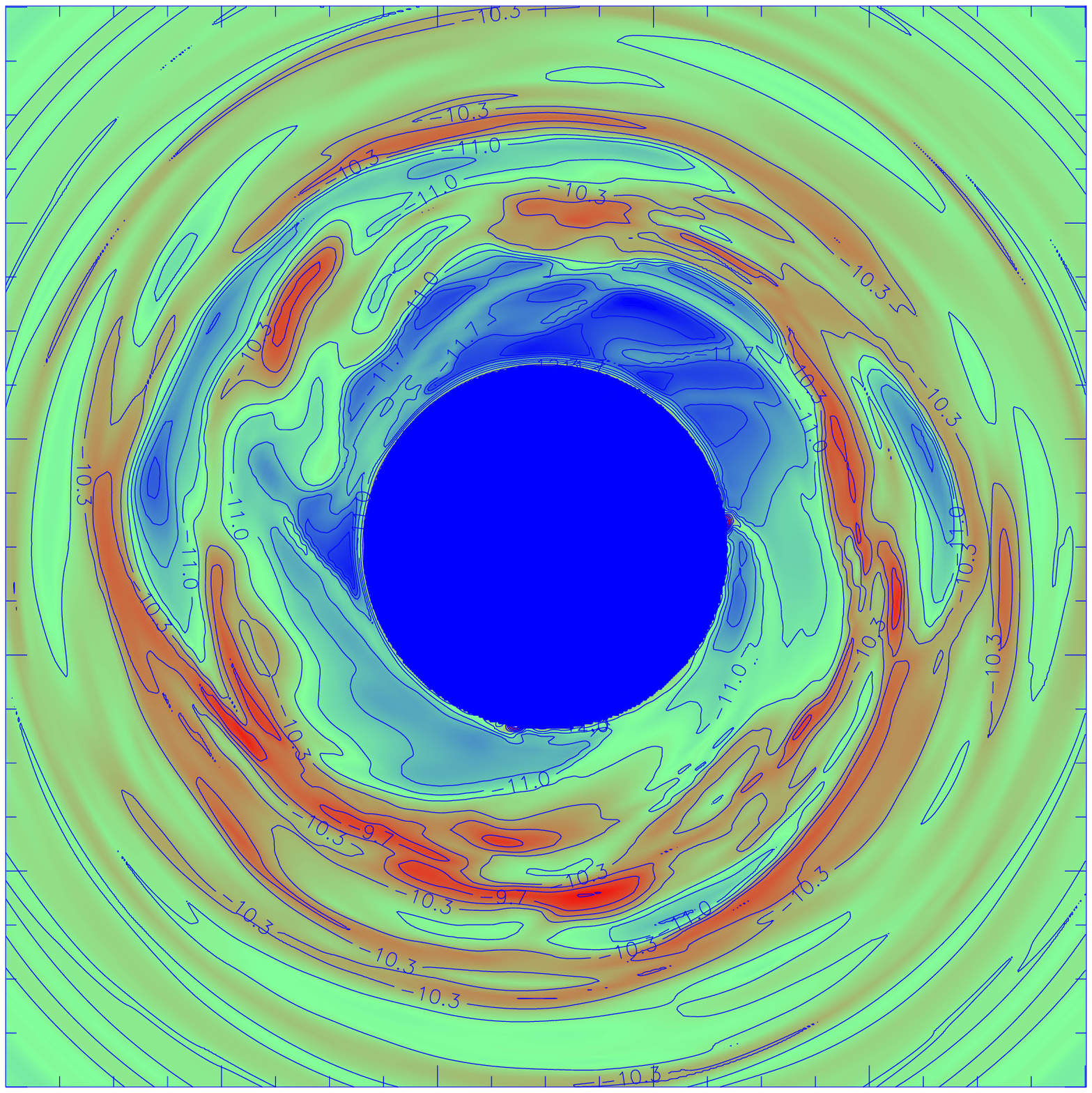}
\caption{Equatorial log density for model A after 129 yr of evolution. 
Colors span a rainbow running from blue (low density) to red (high density). 
Contours are spaced by factors of $\sim 2$ in density. Region shown is
24 AU wide; inner region (blue) is 4 AU in radius.}
\end{figure}

\begin{figure}
\plotone{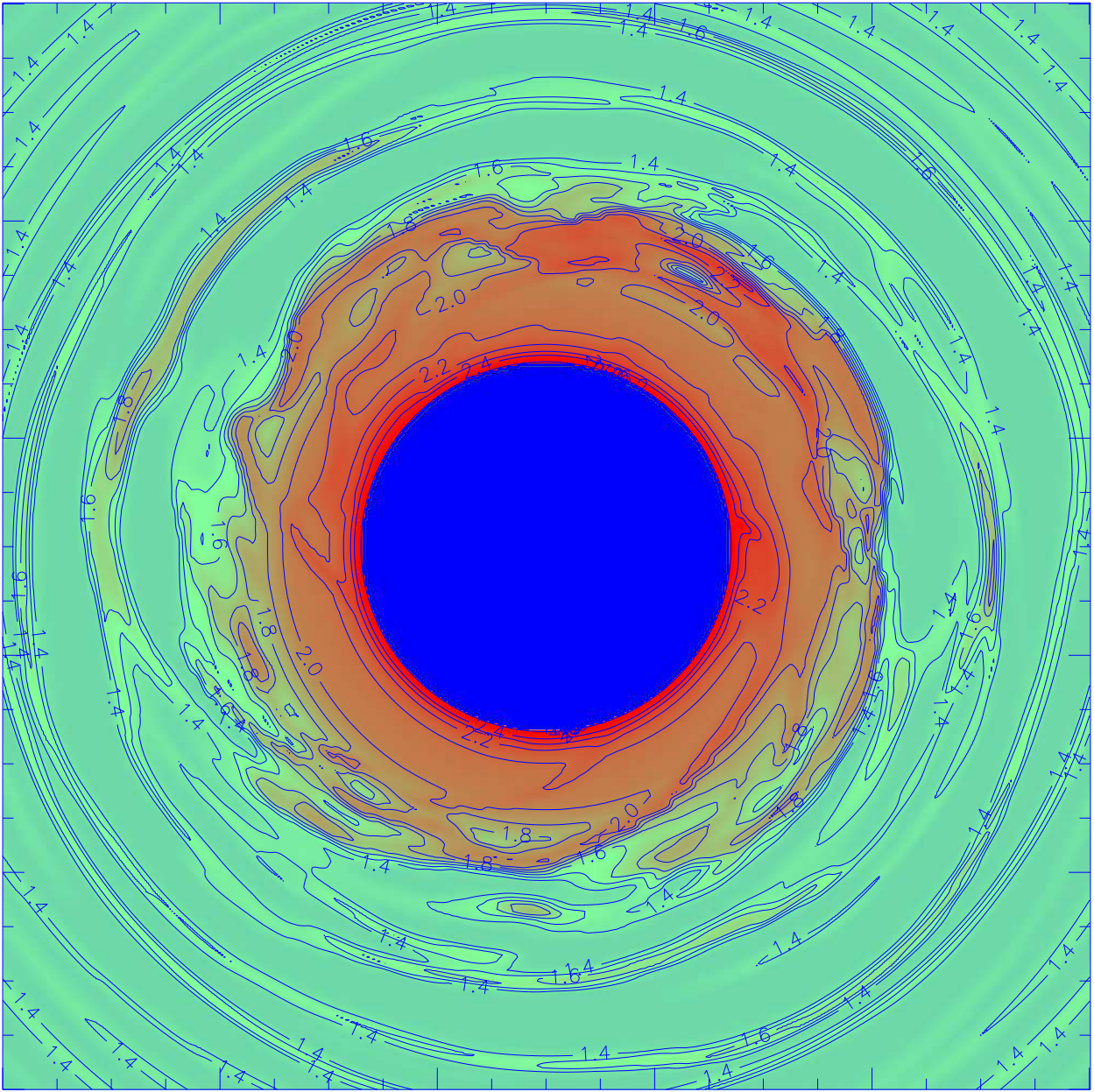}
\caption{Equatorial log temperature for model A after 129 yr of evolution,
plotted as in Figure 1, except that the contours are spaced by factors of 
$\sim 1.3$ in temperature.}
\end{figure}

\begin{figure}
\vspace{-2.0in}
\plotone{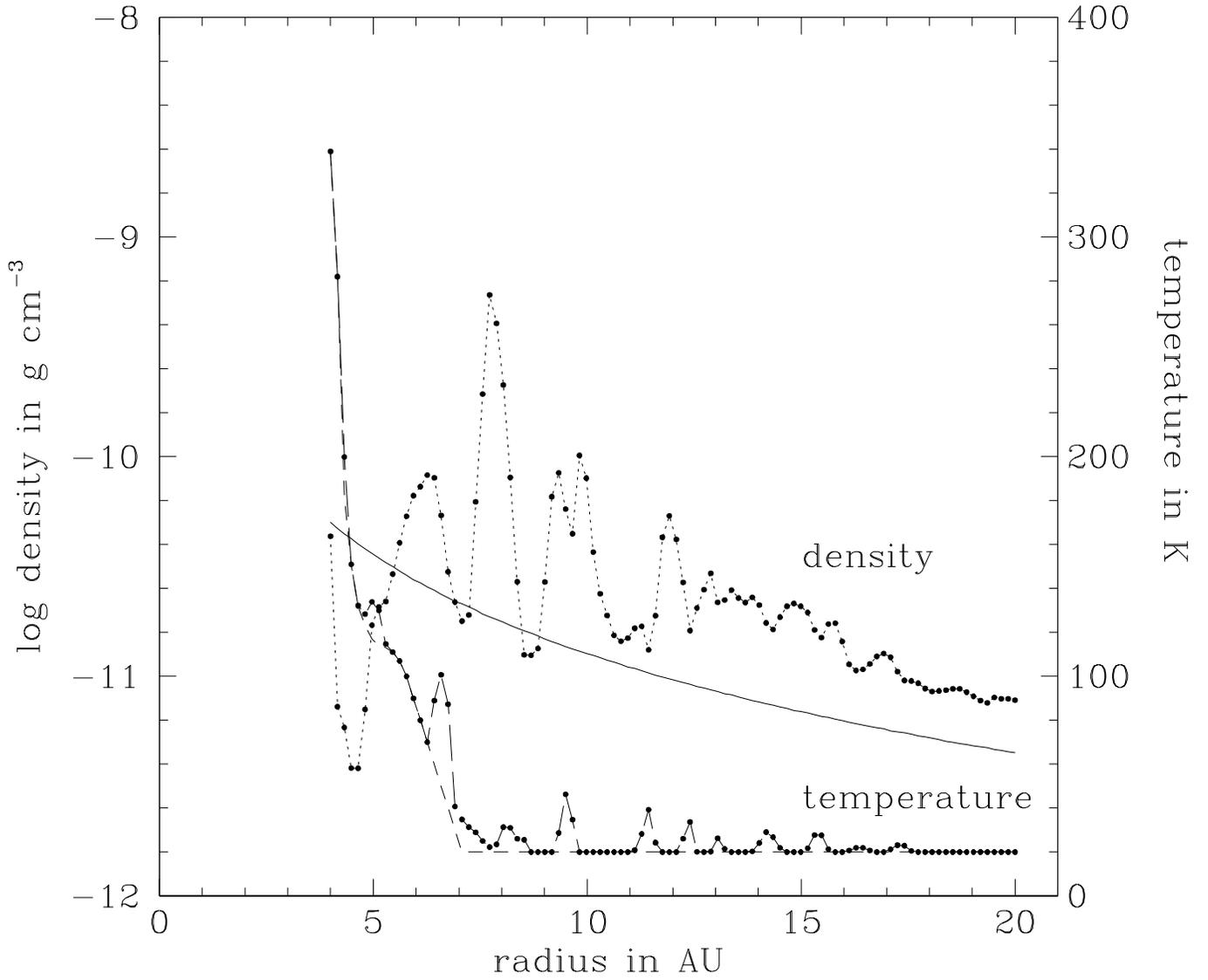}
\caption{Initial log density (solid line) and temperature (dashed line) for 
model A, as well as after 129 yr of evolution (points), for an azimuthal 
profile in the equatorial plane that passes through the clump in Figure 1
at $\sim$ 6 o'clock.}
\end{figure}

\begin{figure}
\plotone{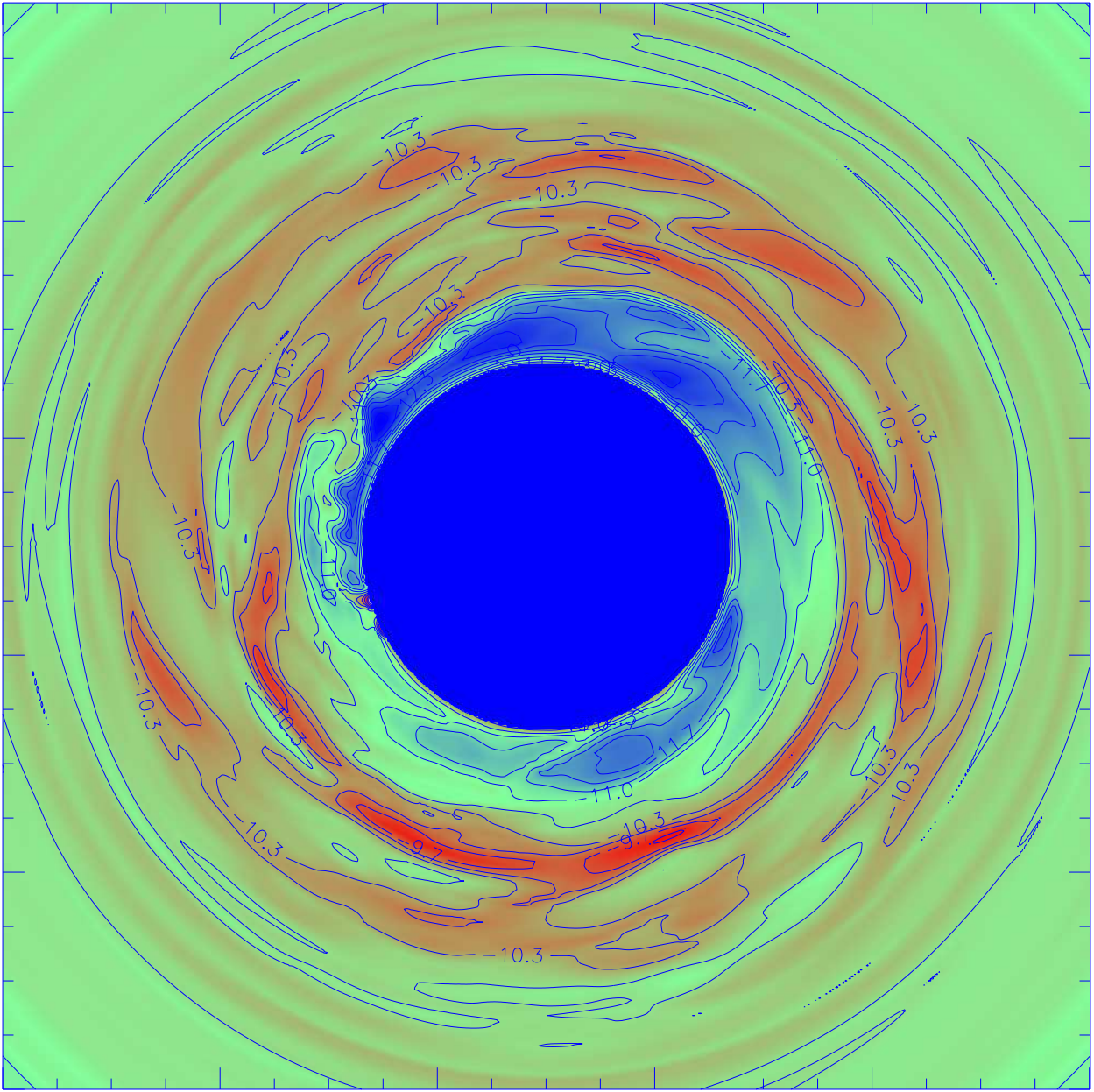}
\caption{Equatorial log density for model B after 119 yr of evolution,
plotted as in Figure 1.}
\end{figure}

\begin{figure}
\plotone{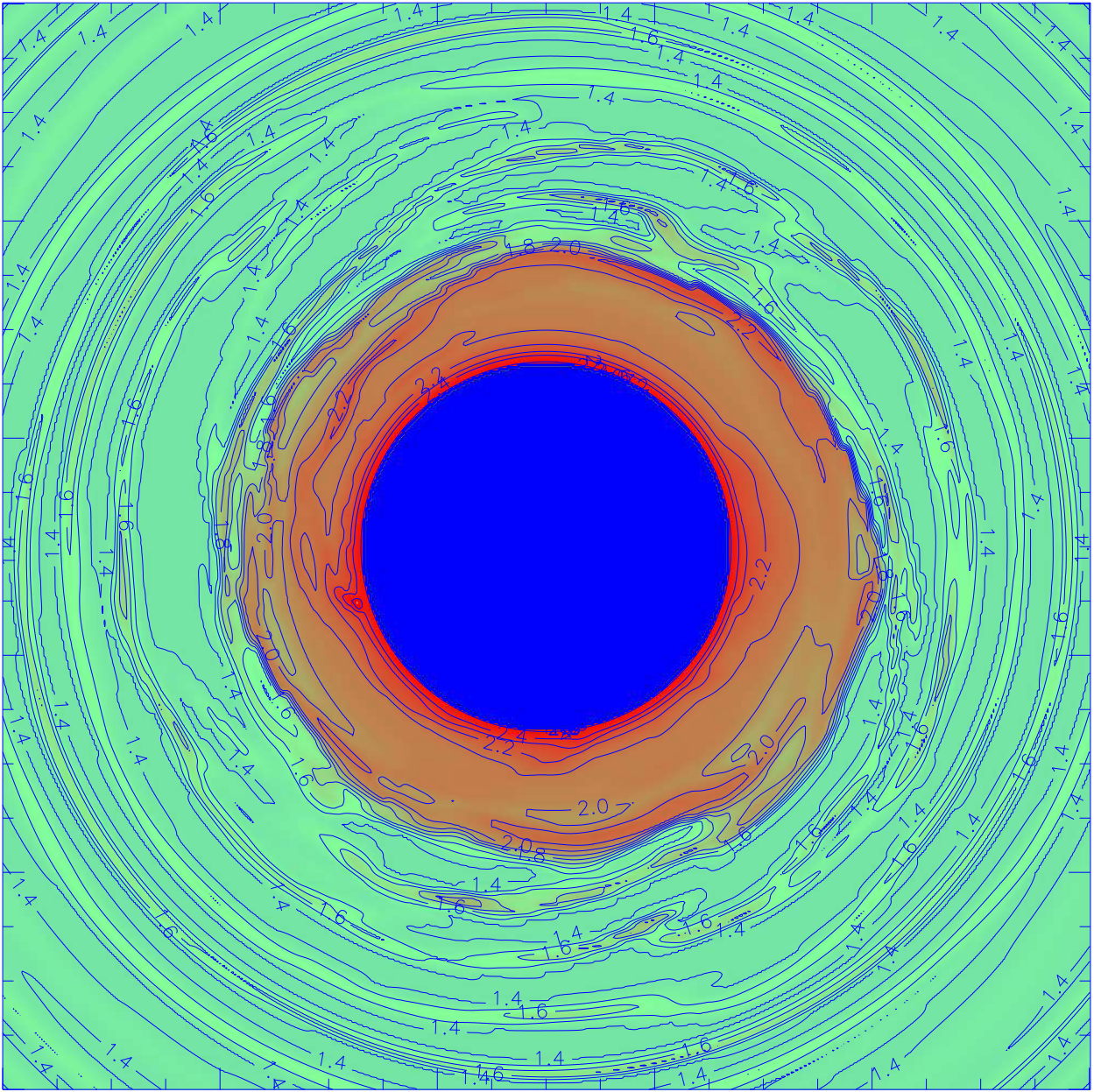}
\caption{Equatorial log temperature for model B after 119 yr of evolution,
plotted as in Figure 2.}
\end{figure}

\end{document}